\documentclass[times]{article}

\usepackage{graphics}     
\usepackage{graphicx}     
\usepackage{longtable}     
\usepackage{url}          
\usepackage{bm}           
\usepackage{textpos}
\usepackage{amsfonts,amsmath,amssymb,mathrsfs,bm,amsthm}
\usepackage{float}
\usepackage{booktabs}
\usepackage{array}
\usepackage{tabu}
\usepackage{dcolumn}
\usepackage{rotating}

\newcommand{\RN}[1]{%
	\textup{\uppercase\expandafter{\romannumeral#1}}%
}

\usepackage{nicefrac}
\usepackage{amsthm}
\usepackage{color}
\usepackage{cancel}
\usepackage{appendix}
\usepackage{authblk}

\newcommand{\appropto}{\mathrel{\vcenter{
			\offinterlineskip\halign{\hfil$##$\cr
				\propto\cr\noalign{\kern2pt}\sim\cr\noalign{\kern-2pt}}}}}
\renewcommand{\v}[1]{\boldsymbol{#1}}		%bold-math for vectors

\usepackage{enumitem,amssymb}
\begin{document}

 \title{Direct limits on the interaction of antiprotons with axion-like dark matter}
%NM-BC: The title is limited to 10 words (or 90 characters)
%% \shorttitle{}
%% Template for a preprint Letter or Article for submission
%% to the journal Nature.
%% Written by Peter Czoschke, 26 February 2004
%%

%% make sure you have the nature.cls and naturemag.bst files where
%% LaTeX can find them

\author[1]{C. Smorra}
\author[2,3]{Y. V. Stadnik}
\author[1,4]{P. E. Blessing}
\author[1,5]{M. Bohman}
\author[1,6]{M. J. Borchert}
\author[1,7]{J. A. Devlin}
\author[1,5,7]{S. Erlewein}
\author[1,5]{J. A. Harrington}
\author[1,8,$\dagger$]{T. Higuchi}
\author[1,5]{A. Mooser}
\author[1,9]{G. Schneider}
\author[1,5]{M. Wiesinger}
\author[1,7]{E. Wursten}
\author[5]{K. Blaum}
\author[8]{Y. Matsuda}
\author[5,10]{C. Ospelkaus}
\author[4]{W. Quint}
\author[2,9]{J. Walz}
\author[1]{Y. Yamazaki}
\author[2]{D. Budker}
\author[1]{S. Ulmer}

\affil[1]{RIKEN, Ulmer Fundamental Symmetries Laboratory, 2-1 Hirosawa, Wako, Saitama, 351-0198, Japan}
\affil[2]{Helmholtz-Institut Mainz, Johannes Gutenberg-Universit{\"a}t, Staudingerweg 18, D-55128 Mainz, Germany}
\affil[3]{Kavli Institute for the Physics and Mathematics of the Universe (WPI), University of Tokyo, 5-1-5 Kashiwanoha, Kashiwa, Chiba 277-8583, Japan}
\affil[4]{GSI-Helmholtzzentrum f{\"u}r Schwerionenforschung GmbH, Planckstrasse 1, D-64291 Darmstadt, Germany}
\affil[5]{Max-Planck-Institut f{\"u}r Kernphysik, Saupfercheckweg 1, D-69117, Heidelberg, Germany}
\affil[6]{Institut f{\"u}r Quantenoptik, Leibniz Universit{\"a}t, Welfengarten 1, D-30167 Hannover, Germany}
\affil[7]{CERN, Esplanade des Particules 1, 1217 Meyrin, Switzerland}
\affil[8]{Graduate School of Arts and Sciences, University of Tokyo, 3-8-1 Komaba, Meguro, Tokyo 153-0041, Japan}
\affil[9]{Institut f{\"u}r Physik, Johannes Gutenberg-Universit{\"a}t, Staudinger Weg 7, D-55099 Mainz, Germany}
\affil[10]{Physikalisch-Technische Bundesanstalt, Bundesallee 100, D-38116 Braunschweig, Germany}

\newcommand{\Addresses}{{% additional braces for segregating \footnotesize
  \bigskip
  \footnotesize

  $\dagger$ Present affliation: Research Center for Nuclear Physics, Osaka University, 10-1 Mihogaoka, Ibaraki, Osaka 567-0047, Japan
}}

\bibliographystyle{naturemag}

\maketitle

\begin{abstract}
Astrophysical observations indicate that there is roughly five times more dark matter in the Universe than ordinary baryonic matter \cite{DM-Review}, with an even larger amount of the Universe's energy content due to dark energy \cite{DE-Review}. 
So far, the microscopic properties of these dark components have remained shrouded in mystery. In addition, even the five percent of ordinary matter in our Universe has yet to be understood, since the Standard Model of particle physics lacks any consistent explanation for the predominance of matter over antimatter \cite{BAU-review}. 
Inspired by these central problems of modern physics, we present here a direct search for interactions of antimatter with dark matter, and place direct constraints on the interaction of ultra-light axion-like particles --- one of the dark-matter candidates --- and antiprotons. 
If antiprotons exhibit a stronger coupling to these dark-matter particles than protons, such a CPT-odd coupling could provide a link between dark matter and the baryon asymmetry in the Universe. 
We analyse spin-flip resonance data acquired with a single antiproton in a Penning trap \cite{SmorraNature} in the frequency domain to search for spin-precession effects from ultra-light axions with a characteristic frequency governed by the mass of the underlying particle. 
Our analysis constrains the axion-antiproton interaction parameter $f_a/C_{\overline{p}}$ to values greater than $0.1$ to $0.6$~GeV in the mass range from $2 \times 10^{-23}$ to $4 \times 10^{-17}\,$eV/$c^2$, improving over astrophysical antiproton bounds by up to five orders of magnitude. 
In addition, we derive limits on six combinations of previously unconstrained Lorentz-violating and CPT-violating terms of the non-minimal Standard Model Extension \cite{Ding2016}. 
	\end{abstract}

A variety of experiments are aiming for the detection of axions and axion-like particles to identify the microscopic nature of dark matter \cite{NPAM-Review,AxionReview}.
Axions are light spinless bosons ($m_a \ll 1~\textrm{eV}/c^2$) originally proposed to resolve the strong CP problem of quantum chromodynamics \cite{Kim2010Review}, and later identified as excellent dark-matter candidates. 
Although limits have been placed on their interaction strengths with photons, electrons, gluons and nucleons \cite{AxionReview,Stadnik2018Review}, direct information on the interaction strength with antimatter is lacking. 
The interactions in the Standard Model have equal couplings to conjugate fermion/antifermion pairs, since the combined charge-, parity- and time-reversal (CPT) invariance is embedded as a fundamental symmetry in the Standard Model. 
CPT invariance has been tested with high sensitivity in recent precision measurements on antihydrogen, antiprotonic helium, and antiprotons \cite{SmorraNature, JerryAntiproton, ALPHA, Masaki, UlmerNature2015, SchneiderScience2017}, and so far no indications for a violation have been found. 
In contrast, the non-observation of primordial antimatter and the matter excess in our Universe are a tremendous challenge for the Standard Model, since the tiny amount of CP-violation contained in the Standard Model is insufficient to reproduce the matter content by more than eight orders of magnitude \cite{BAU-review}. 
However, the discovery of an asymmetric coupling of dark-matter particles to fermions and antifermions may provide an important clue to improve our understanding of dark matter and the baryon asymmetry. Such an asymmetric coupling may in principle arise for axion-like particles if the underlying theory is non-local \cite{Greenberg}, and we test for possible signatures in the spin transitions of a single antiproton. \\

The canonical axion and axion-like particles (collectively referred to as ``axions'' below) can be hypothetically produced in the early Universe by non-thermal mechanisms, such as ``vacuum misalignment'' \cite{Marsh2015Review}. Subsequently, they form a coherently oscillating classical field:~$a \approx a_0 \cos(\omega_a t)$, where the angular frequency is given by $\omega_a \approx m_a c^2 / \hbar$. 
Here, $m_a$ is the axion mass, $c$ the speed of light and $\hbar$ the reduced Planck constant. The axion field carries the energy density $\rho_a \approx m_a^2 a_0^2 /2$, which may comprise the entire local cold dark matter energy density $\rho_\textrm{DM}^\textrm{local} \approx 0.4~\textrm{GeV/cm}^3$ \cite{Catena2010}. 
Assuming that axions are the main part of the observed dark matter, a lower mass bound of $m_a \gtrsim 10^{-22}$ eV is imposed by the requirement that the reduced axion de Broglie wavelength does not exceed the dark-matter halo size of the smallest dwarf galaxies ($\sim 1\,$kpc). \\

Fermions may interact with axions by a so-called derivative interaction causing spin precession \cite{Stadnik2014A}. In the non-relativistic limit, the relevant part of this interaction can be described by the time-dependent Hamiltonian \cite{Stadnik2014A,nEDM2017}:
	\begin{eqnarray}
	\label{NR_Hamiltonian}
	H_{\textrm{int}} (t) \approx \frac{C_{\bar{p}} a_0}{2 f_a} \sin(\omega_a t) ~ \v{\sigma}_{\bar{p}}  \cdot \v{p}_a \, ,
	\end{eqnarray}
where $\v{\sigma}_{\bar{p}}$, $\v{p}_a$ and $C_{\bar{p}}/f_a$ are the Pauli spin-matrix vector of the antiproton, the axion-field momentum vector, and the axion-antiproton interaction parameter, respectively. 
We note that the fundamental theory to produce a CPT-odd operator like in Eq.~(\ref{NR_Hamiltonian}) with $C_{\bar{p}} \ne C_p$ would need to be non-local \cite{Greenberg}. \\

The leading-order shift of the antiproton spin-precession frequency due to the interaction in Eq.~(\ref{NR_Hamiltonian}) is given by: 
\begin{eqnarray}
\label{Axion_antiproton_anomalous_shift}
	\delta \omega_L^{\bar{p}}(t) &\approx \frac{C_{\bar{p}} m_a a_0 \left| \v{v}_a \right|}{f_a} \left[ A \cos(\Omega_\textrm{sid} t + \alpha) + B \right] \sin(\omega_a t) \, ,
\end{eqnarray}
where $\left| \v{v}_a \right| \sim 10^{-3} c$ is the average speed of the galactic axions with respect to the Solar System, 
$\Omega_\textrm{sid} \approx 7.29 \times 10^{-5}~\textrm{s}^{-1}$ is the sidereal angular frequency, and $\alpha \approx -25^\circ$, $A \approx 0.63$, and $B \approx -0.26$ are parameters determined by the orientation of the experiment relative to the galactic axion dark matter flux \cite{NASA_Coordinates} (see the supplementary information). 
We note that the time-dependent perturbation of the antiproton spin-precession frequency in Eq.~(\ref{Axion_antiproton_anomalous_shift}) has three underlying angular frequencies:~$\omega_1 = \omega_a$, $\omega_2 = \omega_a + \Omega_\textrm{sid}$, and $\omega_3 = \left|\omega_a - \Omega_\textrm{sid} \right|$, which for our experiment orientation have approximately evenly-distributed power between the three modes. \\
	
\begin{figure*}[ht!]
    \centerline{\includegraphics[width=15.0cm,keepaspectratio]{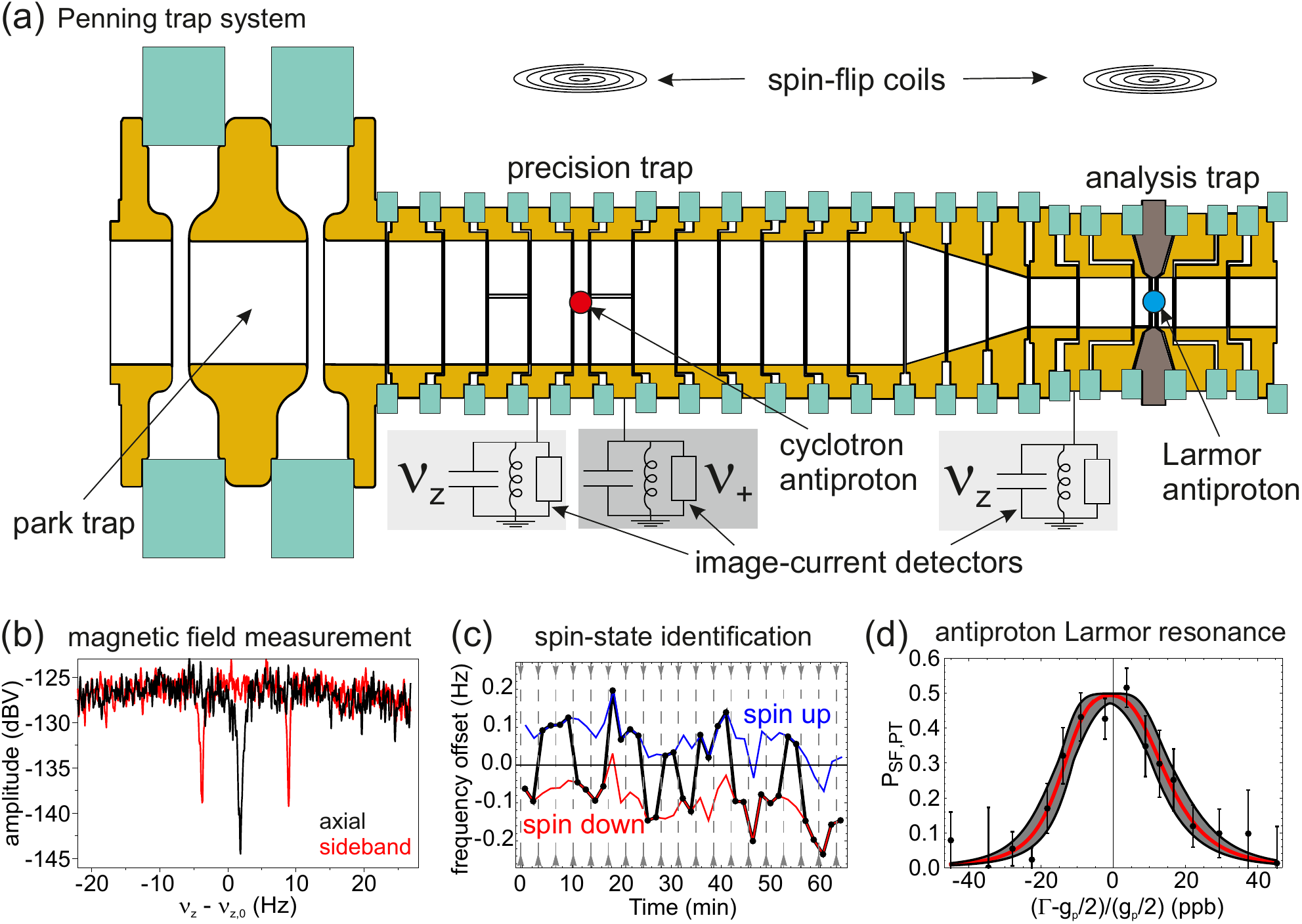}}
     \caption{\textbf{Overview of the antiproton magnetic moment measurement.} 
     (a) The multi-Penning-trap system for the antiproton magnetic moment measurement showing the cyclotron antiproton, the Larmor antiproton, and three Penning traps \cite{SmorraNature}. The trap system consists of a stack of gold-plated copper and CoFe electrodes shown in yellow and brown, respectively, separated by sapphire rings, shown in green. 
     (b) Two FFT spectra of the image-current signal of the cyclotron antiproton for measuring the axial frequency (black curve) and the cyclotron sidebands (red curve). The sideband signal is measured while coupling the axial and cyclotron modes with a quadrupolar radiofrequency drive. The cyclotron frequency $\nu_c$ in the precision trap is extracted from these two spectra \cite{SmorraEPJST2015}.
     (c) A measurement sequence for the identification of the antiproton spin-state in the analysis trap. A series of axial frequency measurements is interleaved by resonant spin-flip drives. The spin state can be assigned with high fidelity by detection of the induced axial frequency shifts  \cite{SmorraPLB2017}.  (d) Larmor resonance of the Larmor antiproton in the precision trap resulting from measuring the spin-flip probability $P_{\mathrm{SF,PT}}$ in the precision trap at the normalized frequency $\Gamma=\nu_{\mathrm{rf}}/\nu_c$. The measurement is referenced to the proton $g$-factor value from 2014: $g_p/2 = 2.792847350(9)$ \cite{MooserNature2014}. The error bars correspond to 1 s.d.~uncertainties.}
     \label{fig:EXP}
\end{figure*}

The experimental data to search for the dark-matter effect were acquired using the Penning-trap system of the BASE collaboration \cite{SmorraEPJST2015} at CERN's Antiproton Decelerator (AD). We have determined the antiproton magnetic moment $\mu_{\overline{p}}$ by measuring the ratio of the antiproton's Larmor frequency $\nu_L$ and the cyclotron frequency $\nu_c$. In a time-averaged measurement, this results directly in a measurement of $\mu_{\overline{p}}$ in units of the nuclear magneton $\mu_N$:
\begin{eqnarray}
\left(\frac{\nu_L} {\nu_c}\right)_{\overline{p}} =  \frac{g_{\overline{p}}}{2} = -\frac{\mu_{\overline{p}}}{\mu_N},
\end{eqnarray} 
 which can be expressed in terms of the antiproton $g$-factor $g_{\overline{p}}$. The relevant part of the apparatus for this measurement is shown in Fig.~\ref{fig:EXP}. We used a multi-trap measurement scheme with two single antiprotons to determine $\mu_{\overline{p}}$ 350-times more precisely than in the best single-trap measurement \cite{HiroNC2017}. Our multi-trap measurement scheme is described in detail in Ref.~\cite{SmorraNature}. \\

 The measurement of $\nu_L/\nu_c$ takes place in the homogeneous precision trap, see Fig.~\ref{fig:EXP} (a). The cyclotron antiproton is used to determine the cyclotron frequency $\nu_c\approx 29.7\,$MHz with a relative precision of a few parts per billion (ppb) \cite{UlmerNature2015} from the spectra of image-current signals such as those shown in Fig.~\ref{fig:EXP} (b). For the measurement of $\nu_L$, the cyclotron antiproton is moved by voltage ramps into the park trap, and the Larmor antiproton is shuttled into the precision trap. We drive spin transitions in the precision trap using an oscillating magnetic field with a frequency $\nu_{\text{rf}}\approx 82.85\,$MHz. 
 To observe these spin transitions, we need to identify the initial and the final spin state of each spin-flip drive in the precision trap. 
 To this end, we transport the Larmor antiproton into the analysis trap and use the continuous Stern-Gerlach effect \cite{DehmeltCSG}, where a strong magnetic curvature of about $3\times 10^{5}\,$T/m$^2$ couples the magnetic moment of the antiproton to its axial motion. 
 As a consequence, spin transitions become observable as an axial-frequency shift of $\Delta\nu_{z,\mathrm{SF}} =\pm 172(8)\,$mHz.
 The spatial separation of the analysis trap from the precision trap strongly reduces line broadening effects from the magnetic inhomogeneity of the analysis trap in the frequency-ratio measurement, which is the key technique to enable precision measurements of $\mu_{\overline{p}}$ at the ppb level.
 The spin-state identification in the analysis trap is performed in a sequence of axial frequency measurements with interleaved resonant spin-flip drives, as shown in Fig.~\ref{fig:EXP} (c). 
 The average fidelity of correctly identifying spin-transitions in the presence of axial frequency fluctuations is $\approx 80\,\%$ \cite{SmorraNature}.  \\
  
  To determine the antiproton $g$-factor, we measured the spin-flip probability $P_{\mathrm{SF,PT}}$ as a function of the frequency ratio $\Gamma= \nu_{\text{rf}} / \nu_{c}$ in the precision trap, which resulted in the antiproton spin-flip resonance shown in Fig.~\ref{fig:EXP} (d). 
  The data consist of 933 spin-flip experiments recorded over 85 days from 05.09.2016 to 27.11.2016. 
  The measurement cycle time of the resonance was not constant mainly due to the statistical nature of the spin-state readout. 
  The median cycle frequency was about $0.38\,$mHz $\approx$ (44$\,$min)$^{-1}$. 
  The spin-flip drive duration was $t_{\textrm{rf}} = $ 8$\,$s with a constant drive amplitude for all data points. 
  The drive frequency was varied in a range of $\pm 45\,$ppb ($\pm 3.7\,$Hz) around the expected Larmor frequency. 
  The time-averaged value of $\mu_{\overline{p}}$ was extracted by matching the lineshape of an incoherent Rabi resonance to the data, which resulted in $g_{\overline{p}}/2 = 2.792\,847\,344\,1(42)$ with a relative uncertainty of 1.5 ppb \cite{SmorraNature}. \\

 The frequency shift in Eq.~(\ref{Axion_antiproton_anomalous_shift}) causes a time-dependent detuning of the drive and the Larmor frequency in each spin-flip experiment. In the following, we consider slow dynamic effects on spin transitions, where $\omega_a/(2\pi) \ll 1 /t_{\textrm{rf}} = 125\,$mHz, so that the variation of the effective Larmor frequency is negligible during the spin-flip drive and does not affect the spin motion on the Bloch sphere. 
 Each spin-flip experiment at the drive time $t$ probes the ``instantaneous value'' of the Larmor frequency  $\omega_L+\delta \omega_L^{\bar{p}}(t)$. \\
%\begin{eqnarray}
%\left(\frac{\omega_L+\delta \omega_L^{\bar{p}}(t)} {\omega_c}\right)_{\overline{p}} =  \frac{\mu_{\overline{p}}}{\mu_N} %\left[1+ \sum_{i=1}^3 b_i \sin\left(\omega_i t + \phi_i \right)\right],
%\label{eq:3}
%\end{eqnarray} 
% where we have expressed $\delta \omega_L^{\bar{p}}(t)$ using the three underlying frequencies $\omega_i$ with their corresponding phases $\phi_i$ and effective amplitudes $b_i$ in dimensionless units.\\

\begin{figure}
    \centerline{\includegraphics[width=8.0cm,keepaspectratio]{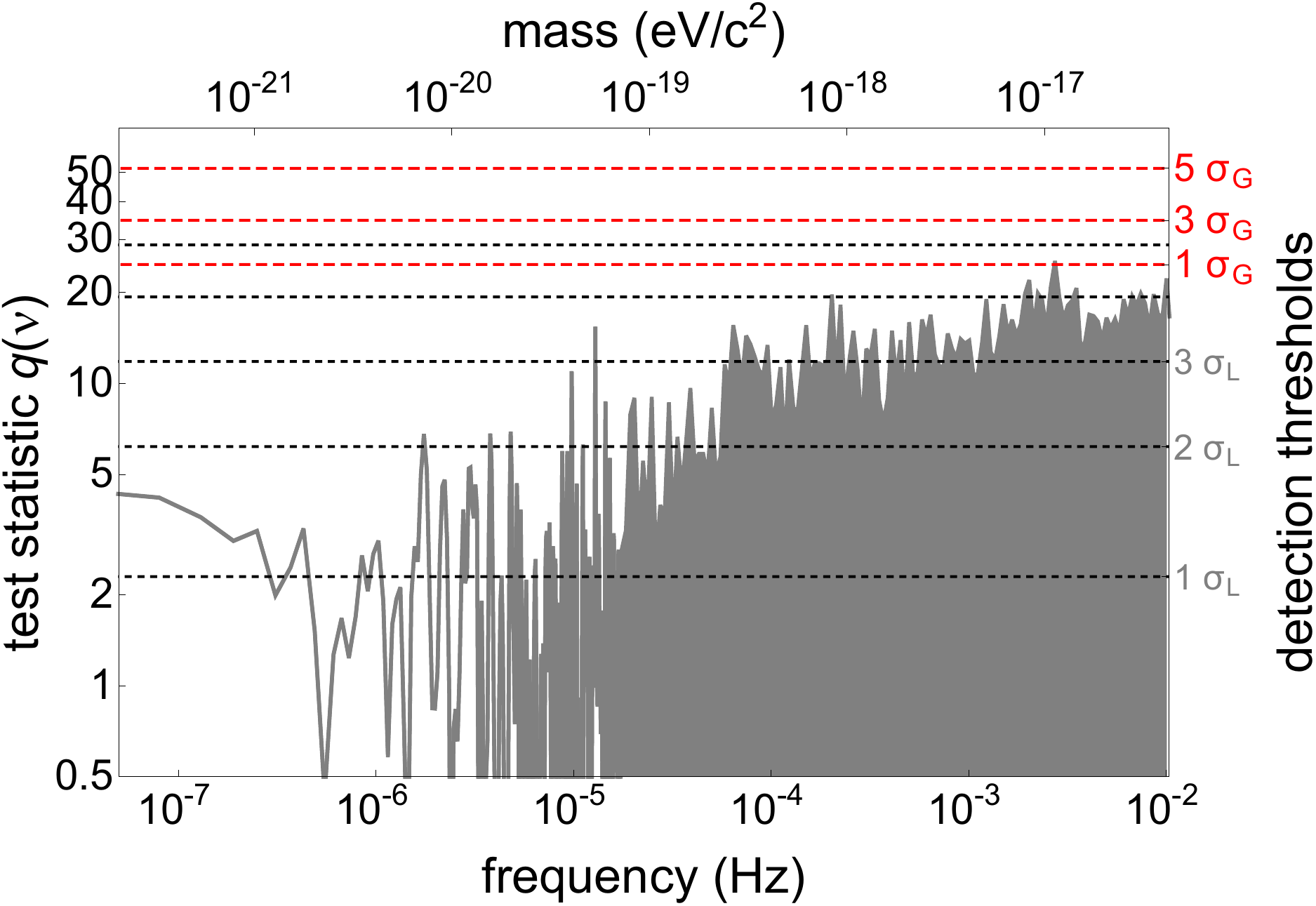}}
     \caption{\textbf{Results of the signal detection.} The test statistic $q(\nu)$ as a function of the frequency $\nu$ is shown as the gray line for the experimental data. The red dashed lines mark the detection thresholds for the global hypothesis test corresponding to 1 (32$\,\%$), 3 (0.27$\,\%$) and 5 standard deviations $\sigma_G$ (5.7$\times10^{-7})$ rejection error for the global test. The black dotted lines show the corresponding statistical significance $\sigma_L$ for a single local test up to 5 $\sigma_L$.} 
     \label{fig:teststat}
\end{figure}

 To conclude whether or not an axion-antiproton coupling is observed, we perform a hypothesis test based on a test statistic $q = - 2 \ln \lambda$, where $\lambda$ denotes the likelihood ratio (see the supplementary information). 
 We compare the zero-hypothesis model $H_0$ with $\delta \omega_L^{\bar{p}}(t)=0$ and extended models $H_{b}(\omega)$, which add an oscillation with frequency $\omega$ to $H_0$, with amplitude $b(\omega) \geq 0$ and phase $\phi(\omega)$ as free parameters.
% To differentiate the zero hypothesis and the extended model, we use the likelihood ratio $q(\omega) = - 2 \ln \left[L_0/L_{b}(\omega)\right]$ as a test statistic \cite{LikelihoodNewPhysics}. $L_0$ and $L_{b}(\omega)$ denote the maximum likelihood of $H_0$ and $H_{b}(\omega)$, respectively, which are maximized in the space of the lineshape parameters for $L_0$. 
% Two additional parameters are included for $L_{b}(\omega)$, namely the amplitude $b(\omega)$ and phase $\phi(\omega)$ introduced above. 
 The test statistic is evaluated for a set of fixed frequencies with a frequency spacing of 60$\,$nHz, which is narrower than the detection bandwidth of our measurement $\approx 1/(T_{\mathrm{meas}})=130\,$nHz. 
 We consider the frequency range $5\,\mathrm{nHz} \leq \omega_i/(2 \pi) \leq 10.49\,\mathrm{mHz}$ in this evaluation and perform a multiple hypothesis test with $N_0= 174\,876$ test frequencies.
 The test statistic as a function of the test frequency is shown in Fig.~\ref{fig:teststat} for the experimental data. 
 To define detection thresholds, we make use of Wilk's theorem to obtain the test-statistic distribution for zero oscillation data, and correct for the look-elsewhere effect (see the supplementary information for details). 
 Based on this, we find that our highest value $q_{\mathrm{max}} = 25.4$ in the entire evaluated frequency range corresponds to a local $p$-value of $p_L = 3 \times 10^{-6}$. 
 This results in a global $p$-value for our multi-hypothesis test of $p_G = 0.254$, which represents the probability that rejecting $H_0$ in favor of any of the alternative models $H_{b}(\omega)$ is wrong. 
 Consequently, we find no significant indication for %an interaction 
 a periodic interaction of the antiproton spin %with an axion field 
 at the present measurement sensitivity, and conclude that our measurement is consistent with the zero hypothesis in the tested frequency range. \\
 
\begin{figure*}
    \centerline{\includegraphics[width=15.0cm,keepaspectratio]{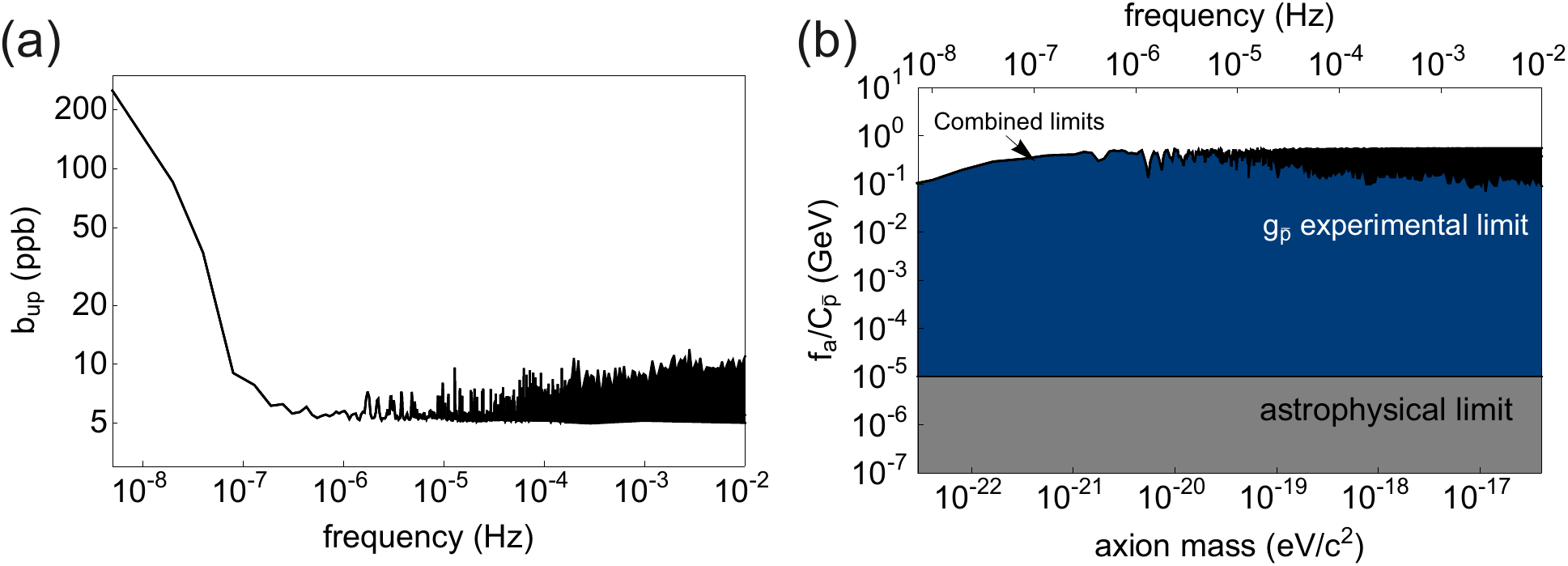}}
     \caption{\textbf{Exclusion limits for the axion-antiproton interaction.} (a) Upper 95$\,\%$ confidence limits on the oscillation amplitude $b_\textrm{up}(\omega)$ of the antiproton Larmor frequency. (b) 95$\,\%$ confidence limits on the axion-antiproton interaction parameter $f_a/C_{\overline{p}}$ as a function of the axion mass. The grey area shows the parameter space excluded by axion emission from antiprotons in SN 1987A. The dark blue area shows the parameter space excluded from our analysis of the antiproton spin-flip data based on the combined limit of the three expected oscillation modes. The black area shows the peak-to-peak difference of the upper experimental exclusion boundary of all tests within a frequency bin.}
     \label{fig:tHist}
\end{figure*}

 To set experimental amplitude limits, we apply the $CL_s$ method \cite{PDG2018} and first extract amplitude limits for single-mode oscillations $b_{\mathrm{up}}(\omega)$ with 95$\,\%$ confidence level. 
% To this end, we estimate the power of the test $\left[1-\beta(b)\right]$ from Monte-Carlo datasets with $b > 0$, and search for the largest amplitude, where $P_L(\omega)/\left[1-\beta(b)\right] > 5\,\%$. 
 The results of $b_{\mathrm{up}}(\omega)$ are shown in Fig.~\ref{fig:tHist} (a).
 In the frequency range $190\,\mathrm{nHz} \leq \omega/(2 \pi) \leq 10\,\mathrm{mHz}$, the mean limit on $b_{\mathrm{up}}$ is 5.5$\,$ppb, which corresponds to an energy resolution of $\sim 2 \times 10^{-24}$ GeV. 
  At lower frequencies $\omega/(2\pi) < 130\,\mathrm{nHz}$, we have sampled only a fraction of an oscillation period. 
  Here, we consider the reduced variation of the Larmor frequency during the measurement and marginalise the quoted limit $b_{\mathrm{up}}(\omega)$ over the starting phase (see the supplementary information).
 To constrain the axion-antiproton coupling coefficient $f_a/C_{\overline{p}}$, we assume that the axion field has a mean energy density equal to the average local dark matter energy density $\rho^{\mathrm{local}}_{\mathrm{DM}} \approx$ 0.4 GeV/cm$^{3}$ during the measurement, and use Eq.~(\ref{Axion_antiproton_anomalous_shift}) to relate $f_a/C_{\overline{p}}$ to the amplitude limits. 
 Since the axion-antiproton coupling would produce almost equal amplitudes at the main frequency $\omega_1$ and the sideband frequencies $\omega_{2,3}$, we place limits on the coupling coefficient considering all three detection modes (see the supplementary information).
 The evaluated limits on the coupling coefficient in the mass range $2 \times 10^{-23}\,$eV/$c^2 < m_a < 4 \times 10^{-17}\,$eV/$c^2$ are shown in Fig.~\ref{fig:tHist} (b).
 The sensitivity of our measurement is mass-independent in the range $m_a \gtrsim 10^{-21}\,$eV/$c^2$, and the amplitude limit is defined by the value of the test statistic at the evaluated mass $q(m_a)$. 
 For $q(m_a) \approx 0$, we obtain $f_a/C_{\overline{p}} > 0.6\,$GeV, which represents the most stringent limitation we can set based on our data.
 In the low-mass range $m_a \lesssim 10^{-21}\,$eV/$c^2$, the amplitude limit on the main frequency $\omega_1$ gets weaker, similar to the behaviour in Fig.~\ref{fig:tHist}(a). 
 The limits in this mass range are dominated by the sideband signals $\omega_{2,3} \approx \Omega_{\mathrm{sid}}$, which remain in the optimal frequency range of our measurement. We also marginalise these limits over the starting phase to account for the possibility of being near a node of the axion field during a measurement (see the supplementary information). 
 These effects lead to less stringent limits for the coupling coefficient for low masses.
 We conclude that we set limits on the axion-antiproton coupling coefficient ranging from $0.1\,$GeV to $0.6\,$GeV in the tested mass range. 
 For comparison, the most precise matter-based laboratory bounds on the axion-nucleon interaction in the same mass range are at the level $f_a/C_N \sim 10^4 - 10^6~\textrm{GeV}$ \cite{nEDM2017,NuclSpinComag}.
 Like in the earlier matter-based studies \cite{nEDM2017,NuclSpinComag}, we do not marginalise our detection limits over possible fluctuations of the axion amplitude $a_0$. 
 We note that preliminary investigations in the recent work \cite{arXiv1905} suggest that, if such amplitude fluctuations are taken into account for sufficiently light axions, then the inferred limits may be weakened by up to an order of magnitude at 95$\%$ C.L.. \\
 
Our laboratory bounds are compared to astrophysical bounds in Fig.~\ref{fig:tHist} (b). In particular, we consider the bremsstrahlung-type axion emission process from antiprotons $\bar{p} + p \to \bar{p} + p + a$ in supernova 1987A, which had a maximum core temperature of $T_\textrm{core} \sim 30~\textrm{MeV}$ and a proton number density of $n_p \sim 5 \times 10^{37}~\textrm{cm}^{-3}$ \cite{Raffelt2008LNP}. 
For an estimate, we treat the supernova medium as being dilute (non-degenerate). 
In thermal equilibrium, this gives the antiproton number density of $n_{\bar{p}} \approx n_p e^{-2 \xi_p / T_\textrm{core}}$, where the proton chemical potential $\xi_p$ is given by $m_p - \xi_p \sim 10~\textrm{MeV}$. 
In the limit of a dilute medium, the axion emission rate from antiprotons scales as $\Gamma_{\bar{p} p \to \bar{p} p a} \propto n_p n_{\bar{p}} \left(C_{\bar{p}}/f_a \right)^2$, whereas the usual axion emission rate from protons scales as $\Gamma_{p p \to p p a} \propto n_p^2 \left(C_p/f_a \right)^2$ \cite{Raffelt2008LNP,Axions_SN1996}. 
Supernova bounds on the axion-proton interaction from the consideration of the effect on the observed neutrino burst duration vary in the range of $f_a/C_p \gtrsim 10^8 - 10^9~\textrm{GeV}$ for $m_a \lesssim T_\textrm{core} \sim 30~\textrm{MeV}$, depending on the specific nuclear physics calculations employed \cite{PDG2018,Raffelt2008LNP}. 
Using the ``middle-ground'' value and rescaling to the axion-antiproton interaction, we obtain the supernova bound $f_a/C_{\bar{p}} \gtrsim 10^{-5}~\textrm{GeV}$ for $m_a \lesssim 30~\textrm{MeV}$, which is up to 5 orders of magnitude weaker than our laboratory bound in the relevant mass range. 
Indirect limits on the axion-antiproton interaction from other astrophysical sources (such as active stars and white dwarves) are even weaker, since their core temperatures are much lower than those reached in supernovae. \\

The non-minimal Standard Model Extension (SME) predicts an apparent oscillation of the antiproton Larmor frequency either at the frequency $\Omega_\textrm{sid}$ or $2\,\Omega_\textrm{sid}$ mediated by Lorentz-violating and in some cases CPT-violating operators added to the Standard Model \cite{Ding2016}. 
With $P_L(\Omega_\textrm{sid}) = 0.336$ and $P_L(2\,\Omega_\textrm{sid}) = 0.328$, we conclude that the zero hypothesis cannot be rejected for these two frequencies, and obtain amplitude limits of $b_{\mathrm{up}}(\Omega_\textrm{sid}) \leq 5.3\,$ppb~and $b_{\mathrm{up}}(2\,\Omega_\textrm{sid}) \leq 5.2\,$ppb with 95$\,\%$ C.L. 
Using these limits and the orientation of our experiment \cite{HiroNC2017}, we constrain six combinations of time-dependent coefficients in the non-minimal SME \cite{Ding2016}:
$ |\tilde{b}^{*X}_p| < 9.7\times 10^{-25}\,\mathrm{GeV}, 
|\tilde{b}^{*Y}_p| < 9.7\times10^{-25}\,\mathrm{GeV}, 
|\tilde{b}^{*XX}_{F,p}-\tilde{b}^{*YY}_{F,p}| < 5.4\times 10^{-9}\,\mathrm{GeV}^{-1},  
|\tilde{b}^{*(XZ)}_{F,p}| < 3.7\times 10^{-9}\,\mathrm{GeV}^{-1}, 
|\tilde{b}^{*(YZ)}_{F,p}| < 3.7\times 10^{-9}\,\mathrm{GeV}^{-1},   
|\tilde{b}^{*(XY)}_{F,p}| < 2.7\times 10^{-9}\,\mathrm{GeV}^{-1}.$
These coefficients are constrained for the first time, since we had only been able to set limits on effects causing a non-zero time-averaged difference of the proton and antiproton magnetic moments \cite{SmorraNature, SchneiderScience2017,HiroNC2017}. \\

In conclusion, our slow-oscillation analysis of the antiproton spin-flip resonance provides the first limits on axion coupling coefficients with an antiparticle probe. Similar searches can be performed for other antiparticles, namely positrons and anti-muons, from frequency-domain analyses of their ($g$-2) measurements \cite{Dehmeltg-2,Muong-2}. 

\Addresses

\textbf{Acknowledgements}\\
We acknowledge technical support by the Antiproton Decelerator group, CERN's cryolab team, and all other CERN groups which provide support to Antiproton Decelerator experiments. We acknowledge discussions with Yunhua Ding about the SME limits, and Achim Schwenk and Kai Hebeler for sharing computing equipment for the Monte-Carlo studies.
We acknowledge financial support by RIKEN, MEXT, the Max-Planck Society, the Max-Planck-RIKEN-PTB Center for Time, Constants and Fundamental Symmetries, the European Union (Marie Skłodowska-Curie grant agreement No 721559), the Humboldt-Program, the CERN fellowship program and the Helmholtz-Gemeinschaft. Y.V.S.~was supported by a Humboldt Research Fellowship from the Alexander von Humboldt Foundation.  D.B.~acknowledges the support by the DFG Reinhart Koselleck project, the ERC Dark-OsT advanced grant (project ID 695405), the Simons and the Heising-Simons Foundations. \\

\textbf{Author contributions} \\
This analysis was triggered by S.U., Y.V.S.~and C.S.. 
C.S.~analysed the experimental data, based on which Y.V.S.~provided the theoretical interpretation and derived the given constraints, which were discussed with D.B.. The manuscript was written by S.U., C.S.,~and Y.V.S., and edited by D.B.. All co-authors discussed and approved the manuscript.\\ 

\textbf{Financial interests} \\
The authors declare no competing financial interests. \\

\textbf{Data availablity} \\
The datasets analyzed for this study will be made available on reasonable request. \\

\textbf{Code availablity} \\
The analysis codes will be made available on reasonable request. \\
 
\textbf{Author information} \\
Reprints and permission information are available at www.nature.com/reprints. 
Correspondence and requests for materials should be addressed to C.S.~\url{Christian.Smorra@cern.ch} or S.U.~\url{Stefan.Ulmer@cern.ch}.\\

\begin{thebibliography}{}

%01-05 Abstract
\bibitem{DM-Review} 
Bertone, G., Hooper, D.~\& Silk, J., 
Particle dark matter:~evidence, candidates and constraints, 
Phys.~Rept.~\textbf{405}, 279 (2005). 

\bibitem{DE-Review} 
Frieman, J.~A., Turner, M.~S.~\& Huterer, D., 
Dark Energy and the Accelerating Universe, 
Ann.~Rev.~Astron.~Astrophys.~\textbf{46}, 385 (2008). 

\bibitem{BAU-review}
Dine, M.~\& Kusenko A.,
Origin of the matter-antimatter asymmetry,
Rev.~Mod.~Phys.~\textbf{76}, 1 (2004).

\bibitem{SmorraNature} 
Smorra, C.~\emph{et al}., 
A parts-per-billion measurement of the antiproton magnetic moment,
Nature \textbf{550}, 371 (2017).

\bibitem{Ding2016} 
Ding, Y.~\& Kostelecky, V.~A.~,
Lorentz-violating spinor electrodynamics and Penning traps,
Phys.~Rev.~D \textbf{94}, 056008 (2016).

%06-15 Second paragraph
\bibitem{NPAM-Review}
Safronova, M.~S., Budker, D., DeMille, D., Kimball, D.~\&, Derevianko, A., 
Search for new physics with atoms and molecules,
Rev. Mod. Phys. \textbf{90}, 025008 (2018). 

\bibitem{AxionReview}
Graham. P.~W.~\emph{et al}.,
Experimental Searches for the Axion and Axion-Like Particles,
Annu. Rev. Nucl. Part. Sci. \textbf{65}, 458 (2015).

\bibitem{Kim2010Review}
Kim, J.~E.~\& Carosi, G.,
Axions and the Strong CP Problem,
Rev.~Mod.~Phys.~\textbf{82}, 557 (2010). 

\bibitem{Stadnik2018Review}
Stadnik, Y.~V.~\& Flambaum, V.~V., 
Searches for New Particles Including Dark Matter with Atomic, Molecular and Optical Systems,
arXiv:1806.03115. 

%10-14 antimatter experiments
\bibitem{JerryAntiproton} 
Gabrielse, G., Khabbaz, A., Hall, D., Heimann, C., Kalinowsky, H. $\&$ Jhe, W.,
Precision mass spectroscopy of the antiproton and proton using simultaneously trapped particles,
\emph{Phys. Rev. Lett.} \textbf{82}, 3198 (1999).

\bibitem{ALPHA} 
Ahmadi, M. \emph{et al.}, 
Chacterization of the 1S-2S transition in antihydrogen
Nature \textbf{557}, 71 (2018).

\bibitem{Masaki} 
Hori, M. \emph{et al.}, 
Buffer-gas cooling of antiprotonic helium to $1.5$ to $1.7\,$K, and antiproton-to-electron mass ratio, 
Science \textbf{354}, 610 (2016).

\bibitem{UlmerNature2015} 
Ulmer, S.~\textit{et al.}, 
High-precision comparison of the antiproton-to-proton charge-to-mass ratio,
Nature \textbf{524}, 196 (2015).

\bibitem{SchneiderScience2017}
Schneider, G.~\emph{et al.},
Double-trap measurement of the proton magnetic moment at 0.3 parts per billion precision,
Science \textbf{358}, 1081 (2017).

\bibitem{Greenberg}
Greenberg, O.~W., 
CPT Violation Implies Violation of Lorentz Invariance, 
Phys.~Rev.~Lett.~\textbf{89}, 231602 (2002). 

%16-17 Axion theory paragraph
\bibitem{Marsh2015Review} 
Marsh, D.~J.~E., 
Axion cosmology, 
Phys.~Rept.~\textbf{643}, 1 (2016). 

\bibitem{Catena2010} 
Catena, R.~\& Ullio, P., 
A novel determination of the local dark matter density, 
J.~Cosmol.~Astropart.~Phys.~\textbf{08} (2010) 004. 

%18-19 p.1 Last paragraph (Eq. 1)
\bibitem{Stadnik2014A} 
Stadnik, Y.~V.~\& Flambaum, V.~V., 
Axion-induced effects in atoms, molecules, and nuclei:~Parity nonconservation, anapole moments, electric dipole moments, and spin-gravity and spin-axion momentum couplings, 
Phys.~Rev.~D~\textbf{89}, 043522 (2014). 

\bibitem{nEDM2017} 
Abel, C.~\emph{et al}., 
Search for Axionlike Dark Matter through Nuclear Spin Precession in Electric and Magnetic Fields, 
Phys.~Rev.~X~\textbf{7}, 041034 (2017). 

%20 p.2 first paragraph (Larmor frequency shift)
\bibitem{NASA_Coordinates} 
NASA LAMBDA -- Tools, 
\url{http://lambda.gsfc.nasa.gov/toolbox/tb_coordconv.cfm}, accessed February 6, 2018. 


%21-23 Fig. 1 Caption 
\bibitem{SmorraEPJST2015} 
Smorra, C.~\emph{et al}., 
BASE - The Baryon Antibaryon Symmetry Experiment,
Eur.~Phys.~J.~Special Topics \textbf{224}, 3055 (2015).

\bibitem{SmorraPLB2017} 
Smorra, C.~\textit{et al.}, 
Observation of individual spin quantum transitions of a single antiproton,
Phys.~Lett.~B \textbf{769}, 1 (2017).

\bibitem{MooserNature2014} 
Mooser, A.~\textit{et al.}, 
Direct high-precision measurement of the magnetic moment of the proton,
Nature \textbf{509}, 596 (2014).

%24 p.2 second paragraph (g-factor measurement)
\bibitem{HiroNC2017} 
Nagahama, H.~\textit{et al.}, 
Sixfold improved single particle measurement of the magnetic moment of the antiproton,
Nat.~Commun.~\textbf{8}, 14084 (2017).

%25, p.3 experiment
\bibitem{DehmeltCSG} 
Dehmelt, H.~\& Ekstr\"om, P.,
Proposed g-2 delta-omegaz experiment on single stored electron or positron,
Bull.~Am.~Phys.~Soc.~\textbf{18}, 727 (1973).

\bibitem{PDG2018}
Tanabashi, M.~\emph{et al}., 
2018 Review of Particle Physics, 
Phys.~Rev.~D~\textbf{98}, 030001 (2018). 

%Inserted PRL on Axion-nucleon coupling
\bibitem{NuclSpinComag}
Wu, T.~\textit{et al.},
Search for Axionlike Dark Matter with a Liquid-State Nuclear Spin Comagnetometer,
Phys.~Rev.~Lett.~\textbf{122}, 191302 (2019).

\bibitem{arXiv1905}
Centers, G.~P.~\emph{et al.},
Stochastic amplitude fluctuations of bosonic dark matter and revised constraints on linear couplings,
arXiv:1905.13650

%30-32 p.3/p.4 astrophysical limits
\bibitem{Raffelt2008LNP} 
Raffelt, G.~G., 
Astrophysical Axion Bounds, 
Lect.~Notes Phys.~\textbf{741}, 51 (2008). 

\bibitem{Axions_SN1996}
Keil, W., Janka, H.-T., Schramm, D.~N., Sigl, G., Turner, M.~S.~\& Ellis, J., 
Fresh look at axions and SN 1987A, 
Phys.~Rev.~D~\textbf{56}, 2419 (1997). 



%33 positron g-2, 34 muon g-2,
\bibitem{Dehmeltg-2}
Van Dyck, R.~S., Schwinberg, P.~B.~\& Dehmelt, H.~G.,
New High-Precision Comparison of Electron and Positron g Factors,
Phys.~Rev.~Lett.~\textbf{59}, 26 (1987).

\bibitem{Muong-2}
Bennett, G.~W.~\emph{et al.},
Search for Lorentz and CPT Violation Effects in Muon Spin Precession,
Phys.~Rev.~Lett.~\textbf{100}, 091602 (2008).

%29 bohman / 30 hannover
%\bibitem{Bohman2018}
%Bohman, M.~\emph{et al.}, 
%Sympathetic cooling of protons and antiprotons with a common endcap Penning trap,
%J.~Mod.~Opt.~\textbf{65}, 568-576 (2018).

%\bibitem{QLEDS2018}
%Meiners, T.~\emph{et al.},
%Towards sympathetic cooling of single (anti-)protons,
%Hyp.~Int.~\textbf{239}, 26 (2018).

\end{thebibliography}
\end{document}

\title{Supplementary Information: Direct limits on the interaction of antiprotons with axion-like dark matter}

\author[1]{C. Smorra}
\author[2,3]{Y. V. Stadnik}
\author[1,4]{P. E. Blessing}
\author[1,5]{M. Bohman} 
\author[1,6]{\newline M. J. Borchert}
\author[1,7]{J. A. Devlin}
\author[1,5,7]{S. Erlewein}
\author[1,5]{J. A. Harrington}
\author[1,8,$\dagger$]{\newline T. Higuchi}
\author[1,5]{A. Mooser}
\author[1,9]{G. Schneider}
\author[1,5]{M. Wiesinger}
\author[1,7]{E. Wursten}
\author[5]{\newline K. Blaum}
\author[8]{Y. Matsuda}
\author[5,10]{C. Ospelkaus}
\author[4]{W. Quint}
\author[2,9]{J. Walz}
\author[1]{\newline Y. Yamazaki}
\author[2]{D. Budker}
\author[1]{S. Ulmer}
\affil[1]{RIKEN, Ulmer Fundamental Symmetries Laboratory, 2-1 Hirosawa, Wako, Saitama, 351-0198, Japan}
\affil[2]{Helmholtz-Institut Mainz, Johannes Gutenberg-Universit{\"a}t, Staudingerweg 18, D-55128 Mainz, Germany}
\affil[3]{Kavli Institute for the Physics and Mathematics of the Universe (WPI), University of Tokyo, 5-1-5 Kashiwanoha, Kashiwa, Chiba 277-8583, Japan}
\affil[4]{GSI-Helmholtzzentrum f{\"u}r Schwerionenforschung GmbH, Planckstrasse 1, D-64291 Darmstadt, Germany}
\affil[5]{Max-Planck-Institut f{\"u}r Kernphysik, Saupfercheckweg 1, D-69117, Heidelberg, Germany}
\affil[6]{Institut f{\"u}r Quantenoptik, Leibniz Universit{\"a}t, Welfengarten 1, D-30167 Hannover, Germany}
\affil[7]{CERN, Esplanade des Particules 1, 1217 Meyrin, Switzerland}
\affil[8]{Graduate School of Arts and Sciences, University of Tokyo, 3-8-1 Komaba, Meguro, Tokyo 153-0041, Japan}
\affil[9]{Institut f{\"u}r Physik, Johannes Gutenberg-Universit{\"a}t, Staudinger Weg 7, D-55099 Mainz, Germany}
\affil[10]{Physikalisch-Technische Bundesanstalt, Bundesallee 100, D-38116 Braunschweig, Germany}

\newcommand{\Addresses}{{% additional braces for segregating \footnotesize
  \bigskip
  \footnotesize

  $\dagger$ Present affliation: Research Center for Nuclear Physics, Osaka University, 10-1 Mihogaoka, Ibaraki, Osaka 567-0047, Japan
}}

\bibliographystyle{naturemag}

\makeatletter
\setcounter{equation}{3}
\makeatother

\maketitle

\section*{Supplementary Discussion} 
\subsection*{Derivation of antiproton spin-precession frequency shift}

In order to derive the leading-order shift of the antiproton spin-precession frequency due to the interaction in Eq.~(1) in the main text, we have to evaluate the geometric factor $\v{\hat{\sigma}}_{\bar{p}} \cdot \v{\hat{p}}_a$, where the hats denote unit vectors. To this end, we need to express both unit vectors in the celestial equatorial coordinate system with the basis $\left\{ \v{\hat{X}}, \v{\hat{Y}}, \v{\hat{Z}} \right\}$, which is a non-rotating coordinate system, where $\v{\hat{Z}}$ coincides with Earth's rotation axis \cite{Kostelecky1999}. \\

The direction of the spin's quantisation axis $\v{\hat{\sigma}}_{\bar{p}}$ is defined in the rotating laboratory frame basis $\left\{ \v{\hat{x}}, \v{\hat{y}}, \v{\hat{z}} \right\}$. We choose $\v{\hat{z}}$ to coincide with the vertical vector pointing upwards at the location of the experiment, $\v{\hat{x}}$ pointing southwards, and $\v{\hat{y}}$ pointing eastwards. The magnetic field of the BASE apparatus is oriented horizontally and rotated counter-clockwise by $\gamma \approx 120^\circ$ from the $\v{\hat{x}}$-axis \cite{HiroNC2017}. 

The transformation between the laboratory and celestial equatorial frames is given by: 
%\begin{widetext}
\begin{equation}
\label{transformation_matrix}
\begin{bmatrix}
\v{\hat{x}} \\ 
\v{\hat{y}} \\ 
\v{\hat{z}} \\
\end{bmatrix}
= 
\begin{bmatrix}
\cos \left( \chi \right) \cos \left( \Omega_\textrm{sid} t \right) & \cos \left( \chi \right) \sin \left( \Omega_\textrm{sid} t \right) & -\sin \left( \chi \right) \\ 
-\sin \left( \Omega_\textrm{sid} t \right) & \cos \left( \Omega_\textrm{sid} t \right) & 0 \\ 
\sin \left( \chi \right) \cos \left( \Omega_\textrm{sid} t \right) & \sin \left( \chi \right) \sin \left( \Omega_\textrm{sid} t \right) & \cos \left( \chi \right) \\
\end{bmatrix}
\begin{bmatrix}
\v{\hat{X}} \\ 
\v{\hat{Y}} \\ 
\v{\hat{Z}} \\
\end{bmatrix}
\, ,
\end{equation}
%\end{widetext}
where $\chi \approx 44^\circ$ is the local colatitude of the experiment, and the ``zero time'' is set by the requirement that the $X$ axis of the celestial equatorial coordinate system has zero right ascension. 
The direction of the quantisation axis in celestial equatorial coordinates is given by: 
\begin{align}
\label{quantisation_axis_CCS}
\v{\hat{\sigma}}_{\bar{p}} &= \cos \left( \gamma \right) \v{\hat{x}} + \sin \left( \gamma \right) \v{\hat{y}} \notag \\
&= \begin{bmatrix}
\cos \left( \gamma \right) \cos \left( \chi \right) \cos \left( \Omega_\textrm{sid} t \right) - \sin \left( \gamma \right) \sin \left( \Omega_\textrm{sid} t \right) \\ 
\cos \left( \gamma \right) \cos \left( \chi \right) \sin \left( \Omega_\textrm{sid} t \right) + \sin \left( \gamma \right) \cos \left( \Omega_\textrm{sid} t \right) \\ 
- \cos \left( \gamma \right) \sin \left( \chi \right) \\
\end{bmatrix}
\, . 
\end{align} \\

To determine the direction of the axion-field momentum vector $\v{\hat{p}}_a$ in celestial equatorial coordinates, we note that in galactic coordinates the Solar System (as it orbits the Galactic centre) moves towards the direction defined by $90^\circ$ longitude and $0^\circ$ latitude. 
Since galactic dark matter is believed to have a Maxwell-Boltzmann-type distribution of velocities with a local average velocity much smaller than the Solar System's velocity, the direction of the axion-field momentum vector in galactic coordinates is hence towards the direction defined by $270^\circ$ longitude and $0^\circ$ latitude. 
Transforming from the galactic coordinate system to the celestial equatorial coordinate system \cite{NASA_Coordinates}, we find that the direction of the axion-field momentum vector is given by: 
\begin{equation}
\label{DM_flux_CCS}
\v{\hat{p}}_{a} = \begin{bmatrix}
\cos \left ( \delta \right) \cos \left( \eta \right) \\ 
\cos \left ( \delta \right) \sin \left( \eta \right) \\ 
\sin \left ( \delta \right) \\
\end{bmatrix}
\, , 
\end{equation}
where $\delta \approx -48^\circ$ and $\eta \approx 138^\circ$ are the declination and right ascension, respectively, of the axion-field momentum relative to the Solar System. 
Using Eqs.~(\ref{quantisation_axis_CCS}) and (\ref{DM_flux_CCS}), we can now compute the geometric factor $\v{\hat{\sigma}}_{\bar{p}} \cdot \v{\hat{p}}_a$: 
\begin{equation}
\label{geom_factor_wind}
\v{\hat{\sigma}}_{\bar{p}} \cdot \v{\hat{p}}_a = A \cos \left( \Omega_\textrm{sid} t + \alpha \right) + B \, ,
\end{equation}
with $\alpha \approx -25^\circ$, $A \approx 0.63$ and $B \approx -0.26$, and hence obtain Eq.~(2) in the main text. 
Further, we can express Eq.~(2) in terms of these three parameters and $\omega_{1,2,3}$: 
\begin{eqnarray}
\label{powerdistr}
    \frac{\delta \omega_L^{\bar{p}}(t) f_a}{C_{\bar{p}} m_a a_0 \left| \v{v}_a \right|}   \approx 
	 B \sin(\omega_1 t) 
	 + \frac{A}{2} \sin(\omega_2 t + \alpha) 
	  - \textrm{sign}(\Omega_\textrm{sid} - \omega_a) \frac{A}{2}	\sin(\omega_3 t + \alpha).
\end{eqnarray}
Consequently, the mode at $\omega_{1}$ carries about 70$\,\%$ of the power relative to those of the sideband modes $\omega_{2,3}$. \\

\subsection*{Limits on the Standard Model Extension Coefficients}
We set limits on the coefficients of the non-minimal SME by expressing the Larmor frequency shift $\delta\omega_L^{\overline{p}}(t)=\delta\omega_a^{\overline{p}}(t)$ in terms of the non-minimal SME coefficients using the Eqs.~(46) and (64) in Ref.~\cite{Ding2016}. Considering the orientation of the magnetic field $B$, we obtain:
\begin{small}
\begin{eqnarray}
\hbar\delta\omega_a^{\overline{p}}(t) = - 2 \tilde{b}^{*3}_p + 2 \tilde{b}^{*33}_{F,p} B = \nonumber \\
2 \tilde{b}^{*X}_p \left[-\cos{(\gamma)} \cos{(\chi)} \cos{(\Omega_{\mathrm{sid}}t)}+\sin{(\gamma)} \sin{(\Omega_{\mathrm{sid}}t)} \right] + \nonumber \\
2 \tilde{b}^{*Y}_p \left[-\cos{(\gamma)} \cos{(\chi)} \sin{(\Omega_{\mathrm{sid}}t)}-\sin{(\gamma)} \cos{(\Omega_{\mathrm{sid}}t)} \right] + \nonumber \\
2 \tilde{b}^{*(XZ)}_{F,p} B \left[-\cos^2{(\gamma)} \sin{(2 \chi)} \cos{(\Omega_{\mathrm{sid}}t)}+\sin{(2 \gamma)} \sin{(\chi)} \sin{(\Omega_{\mathrm{sid}}t)} \right] + \nonumber \\
2 \tilde{b}^{*(YZ)}_{F,p} B\left[-\cos^2{(\gamma)} \sin{(2 \chi)} \sin{(\Omega_{\mathrm{sid}}t)}-\sin{(2 \gamma)} \sin{(\chi)} \cos{(\Omega_{\mathrm{sid}}t)} \right] + \nonumber \\
2 \tilde{b}^{*(XY)}_{F,p} B\left[(\cos^2{(\gamma)} \cos^2{(\chi)} - \sin^2{(\gamma)}) \sin{(2 \Omega_{\mathrm{sid}}t)} + \sin{(2 \gamma)} \cos{(\chi)} \cos{(2 \Omega_{\mathrm{sid}}t)} \right] + \nonumber \\
(\tilde{b}^{*(XX)}_{F,p}-\tilde{b}^{*(YY)}_{F,p}) B \left[(\cos^2{(\gamma)} \cos^2{(\chi)} - \sin^2{(\gamma)}) \cos{(2 \Omega_{\mathrm{sid}}t)} - \sin{(2 \gamma)} \cos{(\chi)} \sin{(2 \Omega_{\mathrm{sid}}t)} \right]
,
\end{eqnarray}
\end{small}
where we have set all combinations of SME coefficients producing a constant frequency shift to zero. We use the single-mode amplitude limit at either $ \Omega_{\mathrm{sid}}$ or $2 \Omega_{\mathrm{sid}}$ to set obtain the limits quoted in the main text on the respective coefficients.

\subsection*{Signal Detection}%\\

In this part, we provide details on the zero-hypothesis tests, where we search for a statistically significant oscillation of the Larmor frequency at the three frequencies $\omega_1 = \omega_a$, $\omega_{2,3}= \left|\omega_a \pm \Omega_{\mathrm{sid}}\right|$ in the experimental data. 
Therefore, we use the test statistic \cite{LikelihoodNewPhysics}:
\begin{eqnarray}
q(\omega)= -2 \ln \lambda(\omega),
\end{eqnarray}
to investigate if our sequence of spin-flip experiments is more likely produced by a time-dependent Larmor frequency given by: 
\begin{eqnarray}
 \omega_L(t) = \omega_L (1+ b \cos\left[\omega t + \phi\right]).
 \label{eq:singlemode}
\end{eqnarray}
The likelihood ratio is given by $\lambda(\omega)=\ln\left[L_0\right]-\ln\left[L_b(\omega)\right]$, where $L_0$ and $L_{b}(\omega)$ denote the maximum likelihood of $H_0$ and $H_{b}(\omega)$, respectively. 
The likelihood is maximized in the manifold of the lineshape parameters for both hypotheses $H_0$ and $H_{b}(\omega)$, and two additional parameters for $H_{b}(\omega)$, namely the amplitude $b(\omega)$ and phase $\phi(\omega)$ introduced above. \\

The likelihood function for the zero-hypothesis test and the parameter exclusion are given by:
\begin{eqnarray}
L = \frac{1}{2^k}\prod_k \left[ 1 - P_k(\mathrm{SF}) - P_{\mathrm{SF},k}
+2 P_k(\mathrm{SF}) P_{\mathrm{SF},k}\right],
\label{eq:lhfunction}
\end{eqnarray}
where $k$ runs over all data points. $P_k(\mathrm{SF})$ is the probability that the spin-flip drive caused a spin transition given the axial frequency shifts measured in the analysis trap preceding and following spin-flip drive $k$. 
Details on evaluating $P_k(\mathrm{SF})$ are given in Refs.~\cite{SmorraNature,SmorraPLB2017}. 
$P_{\mathrm{SF},k}=P_\mathrm{SF}(\Gamma_k, b(\omega), \phi(\omega), \v{a})$ is the spin-flip probability of spin-flip drive $k$ obtained from the lineshape function described in Ref.~\cite{SmorraNature}. 
Here, we replace the Larmor frequency with Eq.~(\ref{eq:singlemode}) to add the time-dependence for the analysis. $\Gamma_k=\nu_{\mathrm{rf},k}/ \left\langle \nu_{c,k,i} \right\rangle_{i=1,...,6}$ is the frequency ratio of the spin-flip drive frequency $\nu_{\mathrm{rf},k}$ divided by the mean of the six corresponding cyclotron frequency measurements $\nu_{c,k,i}$. $\v{a}$ is a vector of nuisance parameters related to the lineshape of the spin-flip resonance, namely the time-averaged antiproton $g$-factor, the Rabi frequency of the spin-flip drive, and the magnitude of magnetic-field fluctuations \cite{SmorraNature}. 
Systematic corrections to $\Gamma_k$ need to be considered to extract $\mu_{\overline{p}}$ and are described in detail in Ref.~\cite{SmorraNature}; however, they are not relevant to constrain the time-dependent effects of interest in this study. \\

Detection thresholds for a local hypothesis test at the frequency $\nu$ are defined by the probability to find data which are less compatible than the observed value $q_{\mathrm{obs}}$ for the experimental data, which are represented by the local $p$-values $p_L(\omega)$. 
To this end, it is necessary to know the test-statistic distribution for the zero-hypothesis data. 
According to Wilk's theorem \cite{Wilks}, we expect that it is a $\chi^2_{f=2}$ distribution with $f=2$ degrees of freedom for zero-hypothesis data evaluated at uncorrelated frequencies $|\nu_1 - \nu_2| \gg 1 / T_\mathrm{meas}$.
We test this assumption by using zero-hypothesis Monte-Carlo datasets to estimate the cumulative density function (CDF) of the test-statistic distribution. The Monte-Carlo datasets are generated using the exact lineshape function described in Ref.~\cite{SmorraNature}. For the evaluation, we approximate the lineshape function for computational efficiency by a Gaussian function $P_{\mathrm{SF}}(\Gamma,C_{\overline{p}},\phi,\v{a})=\mathcal{A}/(\sqrt{2 \pi} \sigma) \exp{\left[-(\Gamma-g_{\overline{p}}/2)^2/(2 \sigma^2)\right]}$, where $g_{\overline{p}}$, $\mathcal{A}$ and $\sigma$ represent the antiproton $g$-factor, an effective amplitude and an effective linewidth, respectively. 
As a result of these simulations, it has been found that this produces a systematic shift of a few ppb in the evaluation of $\mu_{\overline{p}}$, but it preserves the information on time-dependent fluctuations in the data and allows to perform the zero-hypothesis tests for the search of the axion-antiproton interaction.
We evaluated 187 datasets at 320 frequencies with 300$\,$nHz spacing, which  resulted in the test-statistic histogram shown in Supplementary Figure \ref{fig:cdfestimate}, where it is compared to the CDF of the $\chi^2_{f=2}$ distribution. The Monte-Carlo data deviates by 4 standard deviations for $q < 1$, but less than 2 standard deviations for $q > 3$, which is the important region for determining the $P$-values. The reduced $\chi^2$ of our fit is 1.8, and we rely on the $\chi^2_{f=2}$ distribution being a valid approximation of the test-statistic distribution. Consequently, we determine local $p$-values $p_L(\nu)$ according to:
\begin{eqnarray}
p_L(\nu) = 1 - \mathrm{CDF}\left[\chi^2_{f=2}\right](q_{\mathrm{obs}}(\nu)),
\end{eqnarray}
with $\mathrm{CDF}\left[\chi^2_{f=2}\right](q)$ being the CDF of the $\chi^2_{f=2}$-distribution evaluated at $q$. \\

To conclude on the zero-hypothesis test in the entire frequency range, our analysis has to be considered as a multiple hypothesis test. Therefore, we need to refine the detection thresholds considering the number of evaluated tests, the so-called look-elsewhere effect \cite{Elsewhere}. To this end, we evaluate the global $p$-value $p_G$:
\begin{eqnarray}
p_G = 1 - (1 - p_\mathrm{local})^N,
\end{eqnarray}
where $p_\mathrm{local}$ is the smallest value of $p_L(\omega)$ in our test and $N$ is the number of statistically independent tests. 
In our evaluation, we choose a frequency spacing of $\Delta\nu =60\,$nHz $\approx 1/(2 T_\mathrm{meas})$ in our evaluation to avoid missing signals, but consequently the test-statistic values of two neighbouring frequencies with $\Delta\nu < 1 / T_\mathrm{meas}$ are not statistically independent. We determine the correlation factor for the evaluation with 60 nHz spacing by using zero-hypothesis Monte-Carlo datasets. The correlation factor as a function of the frequency difference $\Delta\nu$ is shown in Supplementary Figure \ref{fig:corr}. From this, we extract a correlation factor $\eta = 0.56$ and obtain $N = \eta N_0 =$ 97931. With our lowest local $p$-value of $p_\mathrm{local}=3\times 10^{-6}$, we obtain $p_G=0.254$ as a result.

An alternative approach to determine $p_G$ is to evaluate the same number of test frequencies on several zero-hypothesis datasets, and estimate the distribution of $p_\mathrm{local}$. $p_G$ is obtained by taking the fraction of datasets producing a smaller value of $p_\mathrm{local}$ than the experiment. In our case, this procedure is computationally too expensive due to the large number of test frequencies. Instead, we generate $N$ $\chi^2_{f=2}$-distributed random numbers and determine the distribution of $p_\mathrm{local}$ using this approximation. This results in the global detection thresholds in terms of the test statistic $q$ as shown in Supplementary Table \ref{tab:detthres}. These limits are shown as global detection thresholds in Fig.~2 in the main text. This cross-check procedure results in the same value for $p_G$ as quoted above. \\

We note that our detection analysis tests for individual single-mode oscillations and not explicitly for the model in Eq.~(2) in the main text. We verify below in Monte-Carlo simulations that this approach nevertheless efficiently detects the oscillations described by Eq.~(2) in at least one of the three underlying frequencies. Therefore, based on the non-detection of any oscillation at all, we also conclude that we can reject the alternative hypotheses based on Eq.~(2) with $\omega_a$ in the tested frequency range with $p_G \geq 0.254$. \\

\subsection*{Exclusion limits}%\\
We reject the alternative hypothesis $H_b(\nu)$ against the zero hypothesis $H_0$ with 95$\,\%$ confidence level on the amplitude $b$ using the $CL_s$ statistic: $CL_s (q, b) = \left[1-\beta(q,b)\right]/\left[1-\alpha(q)\right]$ \cite{PDG2018}. 
Here, $\alpha(q)$ and $\beta(q,b)$ denote the CDF of the test-statistic distribution for background signals and signals with amplitude $b$, respectively. \\

For the single-mode detection, we note that $\alpha_{\mathrm{SM}}=\mathrm{CDF}\left[\chi^2_{f=2}\right]$, and determine $\beta_{\mathrm{SM}}(q,b_{\mathrm{SM}})$ by analysing Monte-Carlo datasets where we introduce the time-dependence of the Larmor frequency using Eq.~(\ref{eq:singlemode}) for the Larmor frequency in the lineshape function. 
The time and measurement structure of the Monte-Carlo datasets are identical to the experimental data, but the magnetic-field fluctuations on top of the test-frequency ratios $\Gamma$ are randomised, and the $P_k(\mathrm{SF})$ values for the spin-transition detection, see Eq.~(\ref{eq:lhfunction}), are obtained by randomly redistributing the observed $P_k(\mathrm{SF})$ values of the measurement sequence to the Monte-Carlo data points. \\

We determine the test-statistic distribution on a grid of 12 amplitudes $b_\mathrm{SM}$ and 17 test frequencies $\nu$ including $\Omega_{\mathrm{sid}}/(2\pi)$ and $2 \Omega_{\mathrm{sid}}/(2 \pi)$ to explicitly test the frequencies relevant for the limits on the non-minimal SME coefficients.
Supplementary Figure \ref{fig:powerSM} shows $\beta_{\mathrm{SM}}(q = 5.99,b_{\mathrm{SM}})$, which corresponds to $\alpha_{\mathrm{SM}} \approx 95\,\%$, evaluated over the tested frequency range with about 500 datasets per point.
For frequencies $\nu > 1/T_\mathrm{meas}$ (corresponding to Log$_{10}(\nu/\,$Hz) $\approx$ -7), the power of the test is constant at $\approx$ 95$\,\%$ for amplitudes of 7$\,$ppb. 
Oscillations with lower amplitudes cannot be well distinguished from zero-hypothesis data, since the magnetic-field fluctuations of $\sigma_B/B_0=$3.9(1) ppb observed in cyclotron frequency measurements compete with the detection of interest. 
The single-mode detection efficiency decreases rapidly for frequencies $\nu < 1/T_\mathrm{meas}$, where we sample only a fraction of an oscillation. 
In this frequency range, the change of the Larmor frequency during the measurement decreases for lower frequencies, so that large amplitudes cannot be excluded (further details are discussed below).
Based on $\beta_{\mathrm{SM}}(q,b_{\mathrm{SM}})$ resulting from these Monte-Carlo studies, we determine the 95$\,\%$ C.L. limits on single-mode oscillations shown in Fig.~3(a) in the main text.

To obtain limits on the axion-antiproton coupling coefficient, we need to determine the test-statistic distribution $\beta_i(q,b_a)$ from Monte-Carlo simulations, where we use Eq.~(2) for the time-dependence in the data generation, the index $i$ corresponds to the three detection modes with the frequencies $\omega_i$, and $b_a= C_{\overline{p}} m_a a_0 |\v{v}_a|/(f_a \omega_L)$ expresses the amplitude of the axion field, $a_0$, in units of a relative Larmor frequency shift. 
The distributions $\beta_i(q,b_a)$ are different from $\beta_{\mathrm{SM}}(q,b_{\mathrm{SM}})$, since the detection using a single-mode model needs to cope in this case with the two other modes which effectively act like additional noise sources. 
Supplementary Figure \ref{fig:power} shows the result of evaluating $\beta_i(q = 5.99,b_a)$ for about 700 datasets for each point on a grid of 8 amplitudes and 19 frequencies. 
The presence of the sideband modes $\omega_{2,3}=\left|\omega_{a}\pm\Omega_{\mathrm{sid}}\right|$ perturbs the detection of the main oscillation mode $\omega_1 = \omega_a$, resulting in a test power only slightly above 50$\,\%$ even for large amplitudes.
However, the sideband modes are both efficiently detected with more than 90$\,\%$ power for $A b_a/2  \gtrsim 8\,$ppb. 
The only exception is a narrow window of about 260$\,$nHz around $\omega_a \approx \Omega_{\mathrm{sid}}$ (not resolved in Supplementary Figure \ref{fig:power}), where the lower sideband $\omega_3$ cannot be well detected since its frequency is below 130$\,$nHz. 
However, in this case the power of the test for the other two frequencies $\omega_1$ and $\omega_2$ increases since the lower sideband produces approximately a constant shift of $\omega_L$. \\

For each axion mass $m_a$, we set the most conservative limit on the axion coupling by using the highest test-statistic value of the three test frequencies $\omega_i(m_a)$. 
The power of this test is shown in Supplementary Figure \ref{fig:power}(d) for $q = 5.99$. The test-statistic distribution of the background $\alpha(q)$ in this case is given by the distribution of the maximum of three $\chi^2_{f=2}$-distributed random numbers corresponding to the three different oscillation modes.
This test provides efficient constraints even for $\omega_a < 2 \pi/T_{\mathrm{meas}}$, where we cannot efficiently detect $\omega_1$ or single-mode oscillations, but are sensitive to axion coupling through detection at the sideband frequencies $\omega_{2,3} \approx \Omega_{\mathrm{sid}}$. 
We have explicitly verified the detection down to $\omega_a/(2\pi) = 5\,$nHz in these Monte-Carlo simulations (see also the discussion below). \\

The limits on the axion-induced oscillation amplitude $b_{a}(\omega_i)$ and the coupling parameter are related by:
\begin{eqnarray}
\frac{f_a}{C_{\overline{p}}} > 
\frac{C \sqrt{2\rho_{\mathrm{DM}}\hbar c} |\v{v}_a|}{\omega_L b_{a}(\omega_i)} \approx\frac{2 \,\mathrm{GeV}}{b_{a}(\omega_i)/\mathrm{ppb}},
\end{eqnarray}
where $C$ is either $A/2$ or $|B|$, depending on whether we use the test-statistic value of $\omega_{2,3}$ or $\omega_1$, respectively. The 95$\,\%$ C.L.~amplitude limits for the combined test are shown in Fig.~3(b). Other notations for the axion-nucleon coupling also appear in the literature. The conversion factor to another common notation $g_{aNN}$ \cite{Raffelt2008LNP} is given by $g_{aNN} = C_N m_N / f_a$, while the conversion factor to a less frequently used notation associated with the same symbol $g_{aNN}$ \cite{AxionReview} is given by $g_{aNN} = C_N / (2 f_a)$, where $C_N/f_a$ is the axion-nucleon interaction parameter. \\

\subsection*{Low-frequency limits discussion}
The amplitude limits for the single-mode detection or the main frequency detection $\omega_1 \approx \omega_a$ decreases for low frequencies $\omega_a < 2 \pi / T_{\mathrm{meas}}$, since the measurement time is shorter than the oscillation period. 
If we consider an axion field with a fixed amplitude $a_0$, the maximum difference of the Larmor frequency shift induced during the measurement decreases if $\omega_a \xrightarrow{} 0$. 
Depending on the starting phase, the peak-to-peak difference of the amplitude $\mathcal{B}$ is in the range $B a_0/2 (\varphi/2)^2 < \mathcal{B} < B a_0 \varphi$, where $\varphi = \omega_a T_{\mathrm{meas}}$ is the phase evolution during the measurement. 
Consequently, if we require $\mathcal{B} > \mathcal{B}_{\mathrm{thres}}$ for the detection, $a_0$ needs to increase at least as quickly as $\propto 1/\omega_a$ relative to the high-frequency case as $\omega_a \to 0$ to produce a detectable signal. Consequently, the amplitude limits from the $\omega_1$ detection become less stringent for $\omega_a \xrightarrow{} 0$. 
This effect is present in earlier frequency-domain studies of recent axion experiments \cite{nEDM2017}. \\

The sideband signals at $\omega_{2,3}$ produce amplitude variation on the time scale of a sidereal day, therefore, if the data sample is sufficiently long as in our case, the argument limiting the detection on $\omega_1$ does not apply here. 
However, the sideband signal may also be suppressed if $\varphi \ll 1$ and if the starting phase of the axion field, $\phi$, has a value such that the envelope function in Eq.~(2) is $\sin(\omega_a t) \approx 0$ during the measurement. 
In this case, the sidereal modulation of the amplitude produces a smaller Larmor frequency shift for a fixed  value of $a_0$. If we require that the minimum amplitude modulation for efficient detection is $\mathcal{B}_{\mathrm{thres}}$, then we can detect the axion-coupling in the limit $\omega_a \xrightarrow{} 0$ only under the condition:
\begin{eqnarray}
\mathcal{B}_{\mathrm{thres}} < \frac{A}{2} a_0 |\sin(\phi)|,
\end{eqnarray}
 which results from Eq.~(2) (see also the supplementary material of Ref.~\cite{NuclSpinComag}).
 Since the starting phase of the axion field is unknown, we marginalise the limit over all possible starting phases. Further, if we also average $|\sin(\phi)|$ over the phase evolution during the measurement, we obtain:
\begin{eqnarray}
\begin{split}
\mathcal{B}_{\mathrm{thres}} &< \frac{A}{2} a_0 \frac{1}{2 \pi} \frac{1}{\varphi} \int_{0}^{2 \pi} d\phi \int_{\phi}^{\phi+\varphi} d\gamma |\sin(\gamma)| \\ &\approx 0.636\,\frac{A}{2} a_0,
\end{split}
\end{eqnarray}
which is independent of $\omega_a$ and $T_{\mathrm{meas}}$. Consequently, the detection limit decreases in the transition region $\phi \sim 1$, since the detection is not equally efficient for all starting phases. 
The impact of this effect is shown in Supplementary Figure~\ref{fig:power}, where the power of the sideband detection drops in the range of $A a_0/2 \sim 6\,$ppb from 0.95 to about 0.8 around $\mathrm{Log}_{10}\left[\omega_a/(2\pi)\right] \sim -8$. 
The lower detection power is considered in the limit via the $CL_s$ method and leads to a weaker limit for the axion-antiproton coupling in the transition region, see Fig.~3(b). 
We predict that the detection limit for the sidebands should become independent of $\omega_a$ in the limit $\omega_a \to 0$ for the detection of sideband signals, since the probability of the starting phase being near a node is independent of $\omega_a$.
Our present Monte-Carlo studies, however, do not extend deeply into the low-frequency regime $\varphi \ll 1$, and so our prediction for the low-frequency scaling of the sideband limits needs to be verified in a more extensive analysis.

We also note that it is not possible to determine $\omega_a$ in the low-frequency range in case of a positive detection, since the two sideband signals are within one detection bandwidth of $\approx$ 130 nHz and hence cannot be distinguished. But for our method, it suffices to set limits in the low-frequency range based on the non-detection of signals close to $\Omega_{\mathrm{sid}}$. \\

\newpage

\section*{Supplementary Tables}
\renewcommand{\tablename}{Supplementary Table}
\begin{table*}[h!]
    \centering
    \begin{tabular}{|c|c|c|} \hline
        Statistical significance & $q$ & $p_G$ \\ \hline
        1$\,\sigma_G$ & 24.9 & 0.32 \\
        2$\,\sigma_G$ & 29.1 & 0.046 \\
        3$\,\sigma_G$ & 34.8 & 2.7$\times10^{-3}$ \\
        4$\,\sigma_G$ & 42.3 & 6.3$\times10^{-5}$ \\
        5$\,\sigma_G$ & 50.9 & 5.7$\times10^{-7}$ \\ \hline
    \end{tabular}
    \caption{\textbf{Test-statistic values and global $P$-values for different values of the statistical significance.} }
    \label{tab:detthres}
\end{table*}

\newpage
\section*{Supplementary Figures}
\renewcommand{\figurename}{Supplementary Figure}
\begin{figure*}[h!]
\centering
\includegraphics[width= 0.75\textwidth]{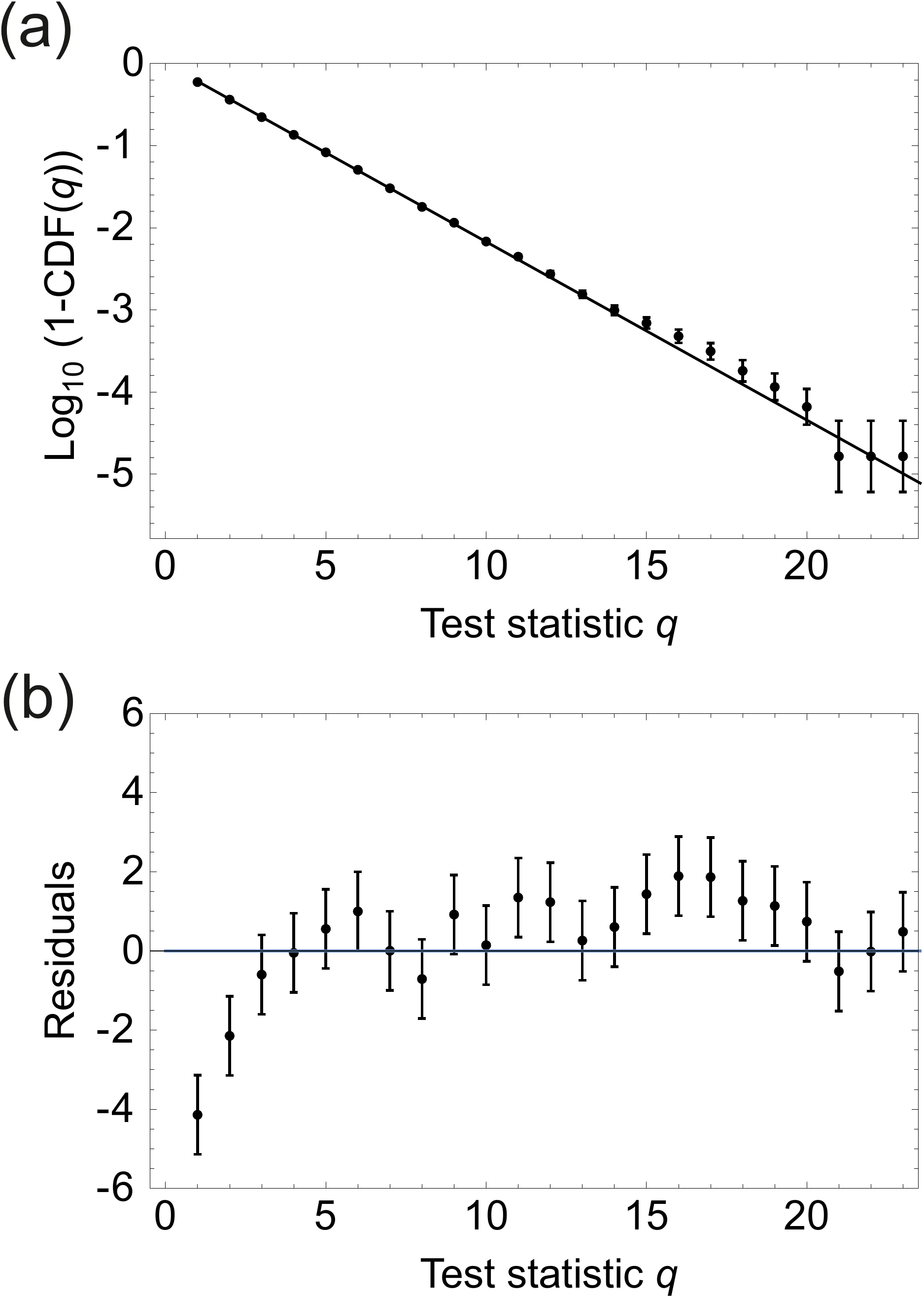}
\caption{\textbf{Monte-Carlo simulation of the test-statistic distribution.} (a) Estimate of the $P$-value as a function of the test statistic $q$. We use zero-hypothesis Monte-Carlo datasets to estimate the cumulative density function of the test-statistic distribution (CDF) and compare it to the CDF of the $\chi^2_{f=2}$ distribution. The reduced $\chi^2$ of the residuals shown in (b) is 1.8. The error bars indicate 1 s.d.~uncertainties.}
\label{fig:cdfestimate}
\end{figure*}

\begin{figure*}[h!]
\centering
\includegraphics[width= 0.85\textwidth]{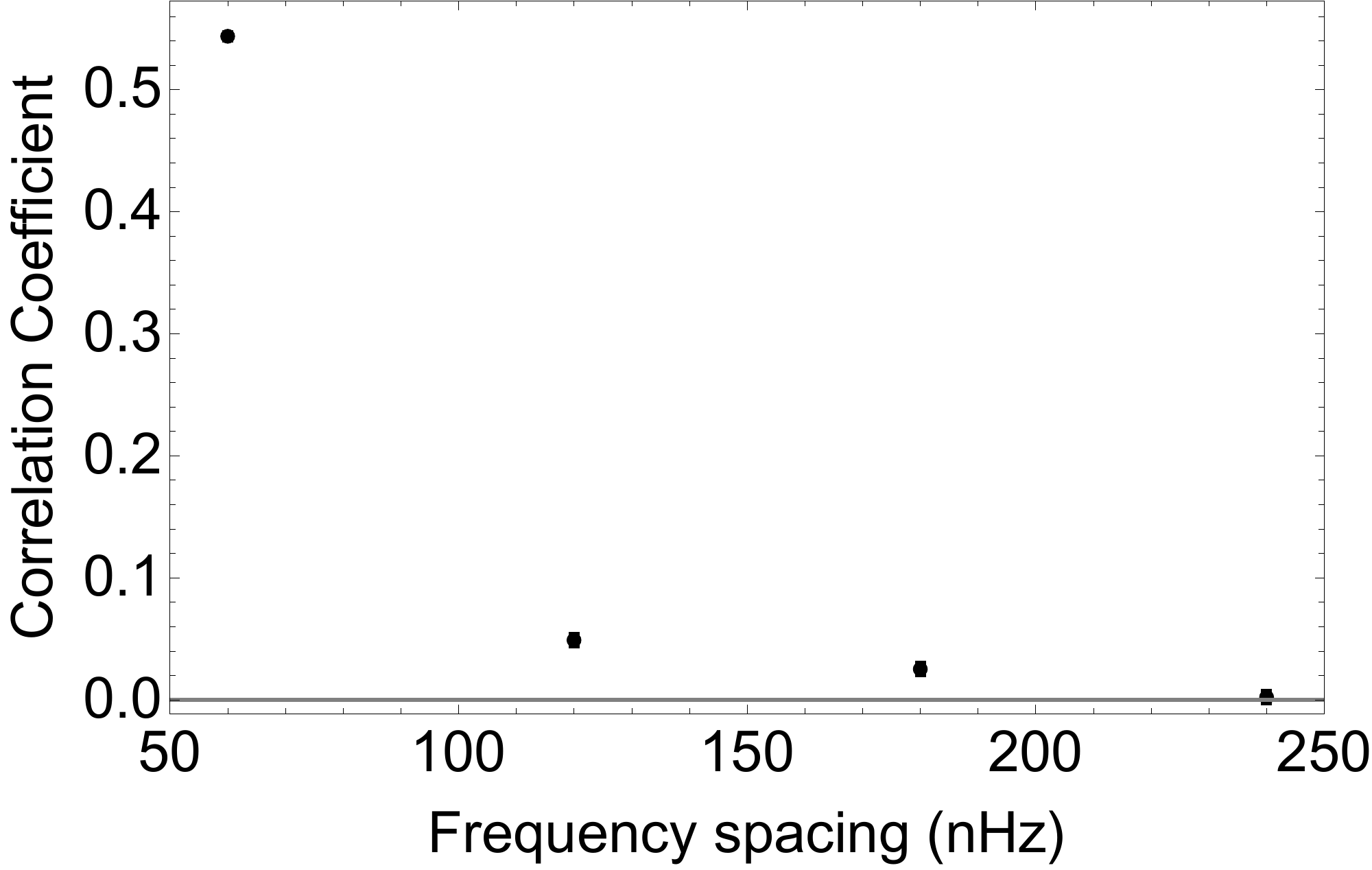}
\caption{\textbf{Test statistic correlation coefficients.} Result of evaluating the correlation coefficients of the test-statistic values $q(\nu$) and $q(\nu+\Delta\nu)$ as a function of the frequency spacing $\Delta\nu$. The error bars indicate 1 s.d.~uncertainties.}
\label{fig:corr}
\end{figure*}

\begin{figure*}[h!]
\centering
\includegraphics[width= 0.98 \textwidth]{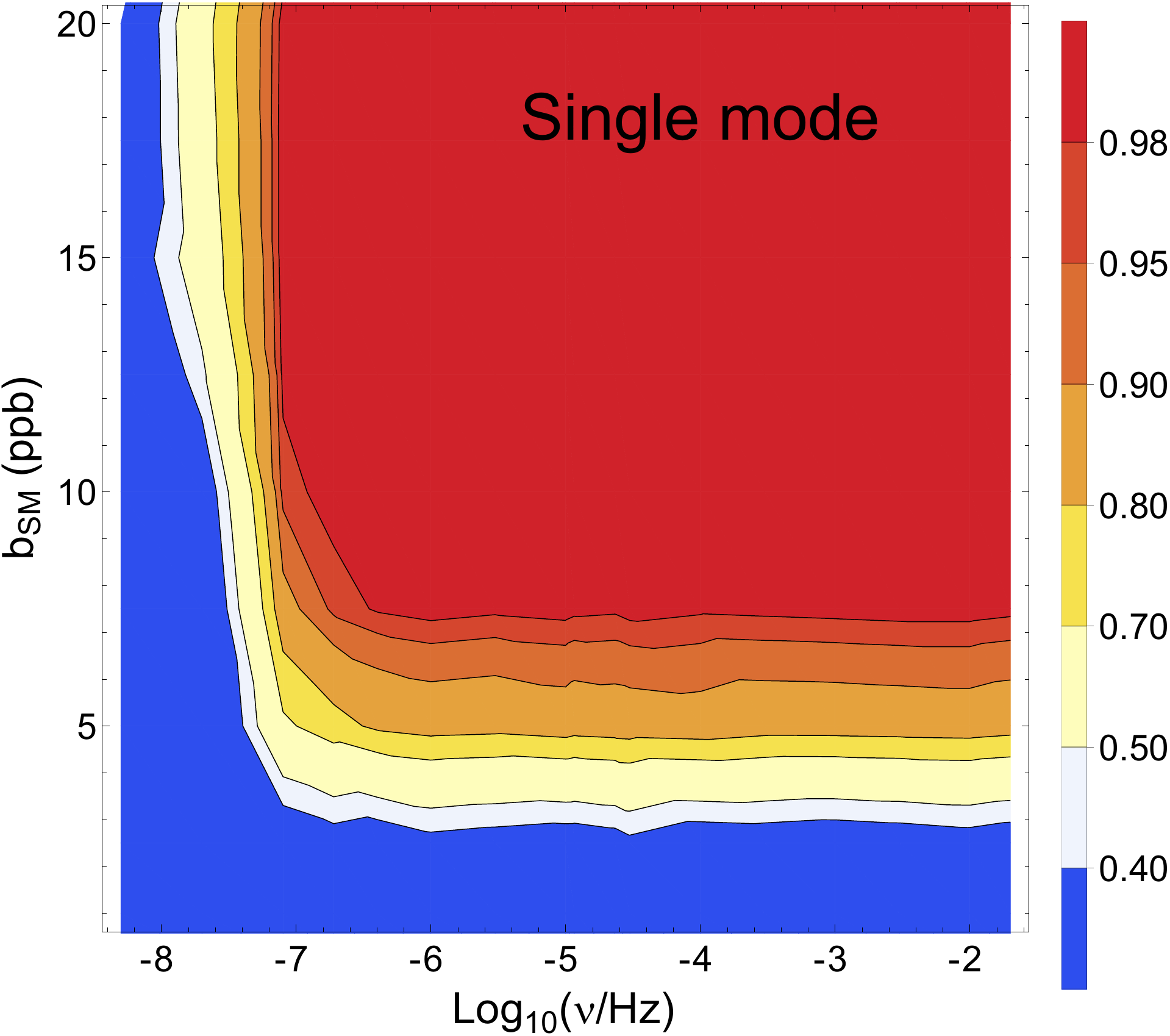}
\caption{\textbf{Power of the hypothesis test for single-mode detection.} The power for detecting a single-mode oscillation with amplitude $b_{\mathrm{SM}}$ is shown at a test-statistic value of $q = 5.99$ ($\alpha = 5\,\%$). This represents the evidence for the alternative model with frequency $\nu$ and amplitude $b_{\mathrm{SM}}$, if the zero hypothesis was rejected with 5$\,\%$ error.}
\label{fig:powerSM}
\end{figure*}

\begin{figure*}[h!]
\centering
\includegraphics[width= 0.95 \textwidth]{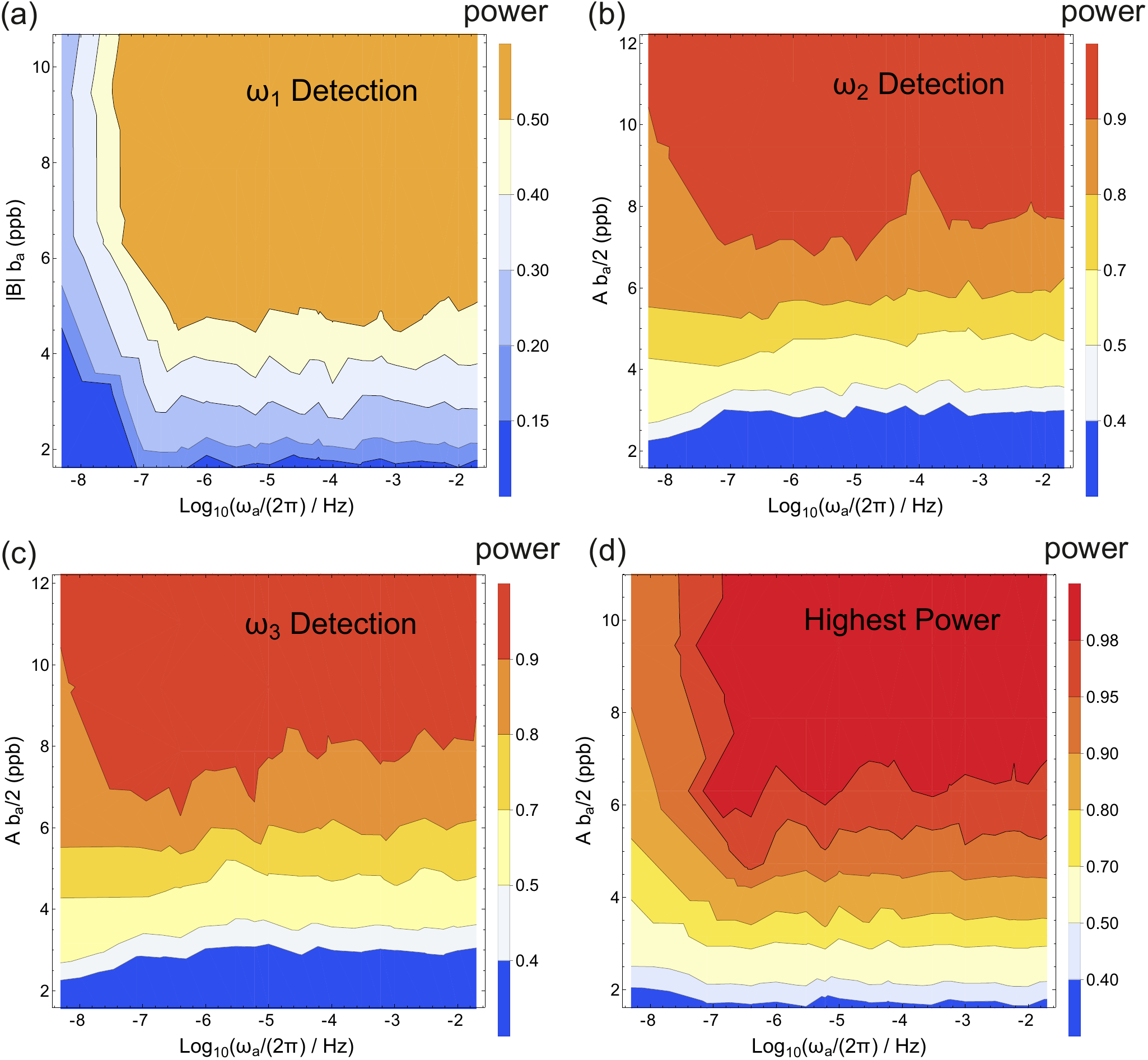}
\caption{\textbf{Power of the hypothesis test for the axion model.} The power for detecting the axion-antiproton coupling at a test-statistic value $q = 5.99$ ($\alpha = 5\,\%$) is shown as a function of $\omega_a/(2\pi)$ and the axion amplitude $b_a$. The four plots are evaluated using different detection criteria: (a) $q(\omega_1)$, (b) $q(\omega_2)$, (c) $q(\omega_3)$ , and (d) for the maximum of these three values.}
\label{fig:power}
\end{figure*}